\documentclass[aps,pre,amsmath,amsfonts,twocolumn,superscriptaddress]{revtex4-1}
\usepackage{graphicx}
\usepackage{dcolumn}
\usepackage{bm}
\usepackage{url}
\usepackage{nicefrac}

\begin{document}


\title{Lassoing saddle splay and the geometrical control of topological defects}



\author{Lisa Tran}
\author{Maxim O. Lavrentovich}
\affiliation{Department of Physics and Astronomy, University of Pennsylvania, Philadelphia PA 19104, United States}
\author{Daniel A. Beller}
\affiliation{School of Engineering and Applied Sciences, Harvard University, Cambridge MA 02138, United States}
\author{Ningwei Li}
\author{Kathleen J. Stebe}
\affiliation{Department of Chemical and Biomolecular Engineering, University of Pennsylvania, Philadelphia, PA 19104}
\author{Randall D. Kamien}
\email[Electronic address: ]{kamien@physics.upenn.edu}
\affiliation{Department of Physics and Astronomy, University of Pennsylvania, Philadelphia PA 19104, United States}


\date{\today}

\begin{abstract}
{Systems with holes, such as colloidal handlebodies and toroidal droplets, have been studied in the nematic liquid crystal (NLC) 4-cyano-4'-pentylbiphenyl (5CB): both point and ring topological defects can occur within each hole and around the system, while conserving the system's overall topological charge. However, what has not been fully appreciated is the ability to manipulate the hole geometry with homeotropic (perpendicular) anchoring conditions to induce complex, saddle-like deformations. We exploit this by creating an array of holes suspended in an NLC cell with oriented planar (parallel) anchoring at the cell boundaries. We study both 5CB and a binary mixture of bicyclohexane derivatives (CCN-47 and CCN-55). Through simulations and experiments, we study how the bulk saddle deformations of each hole interact to create novel defect structures, including an array of disclination lines, reminiscent of those found in liquid crystal blue phases. The line locations are tunable via the NLC elastic constants, the cell geometry, and the size and spacing of holes in the array. This research lays the groundwork for the control of complex elastic deformations of varying length scales via geometrical cues in materials that are renowned in the display industry for their stability and easy manipulability.}
\end{abstract}


\maketitle


The investigation of mechanisms, both chemical and geometrical, to control and manipulate defects in liquid crystals (LCs) is essential for the use of these defects in the hierarchical self-assembly \cite{pp,yada,ska} of photonic and meta-materials \cite{bp-c, phtnrvz}, as well as for studies in low-dimensional topology \cite{ska, bs-tc, utknt, tord, tnk, opal}.  For instance, the disclination line networks characteristic of blue phases \cite{stbp, bplat} have been proposed to organize colloidal inclusions \cite{bp-c, znovbp}. But can similar three-dimensional disclination line networks be designed in the simpler nematic LC? The ubiquitous use of NLCs in the display industry is a testament to their efficacy in applications. Wide-ranging studies on the role of nematic elasticity in designing tailored defect structures have focused primarily on the familiar splay, twist, and bend deformations. Recently, however, there has been a renewed interest in exploiting saddle-splay deformations \cite{tord, zsd,2016rav}. By confining nematics in cells with properly-designed boundary conditions,  we demonstrate an array of controlled, defect-riddled minimum energy states that form as a result of saddle-splay distortions, excitable by the  system's surfaces. 
\section*{Energy Considerations}

We begin with the Frank free energy for a nematic \cite{dgplc, rdkpr}: 
\begin{align}
F&  =  \int \mathrm{d}^3 x \left\{  \frac{K_1}{2}[\mathbf{n} (\nabla \cdot \mathbf{n})]^2+\frac{K_2}{2}[\mathbf{n} \cdot (\nabla \times \mathbf{n})]^2 \right.  \nonumber \\
& \left.   {}+\frac{K_3}{2}[(\mathbf{n} \cdot \nabla)\mathbf{n}]^2-K_{24} \nabla \cdot [(\mathbf{n} \cdot \nabla)\mathbf{n}-(\nabla \cdot \mathbf{n})\mathbf{n} ]\right\}, \label{eq:Frank}
\end{align}

\begin{figure}
\centerline{\includegraphics[width=0.35\textwidth]{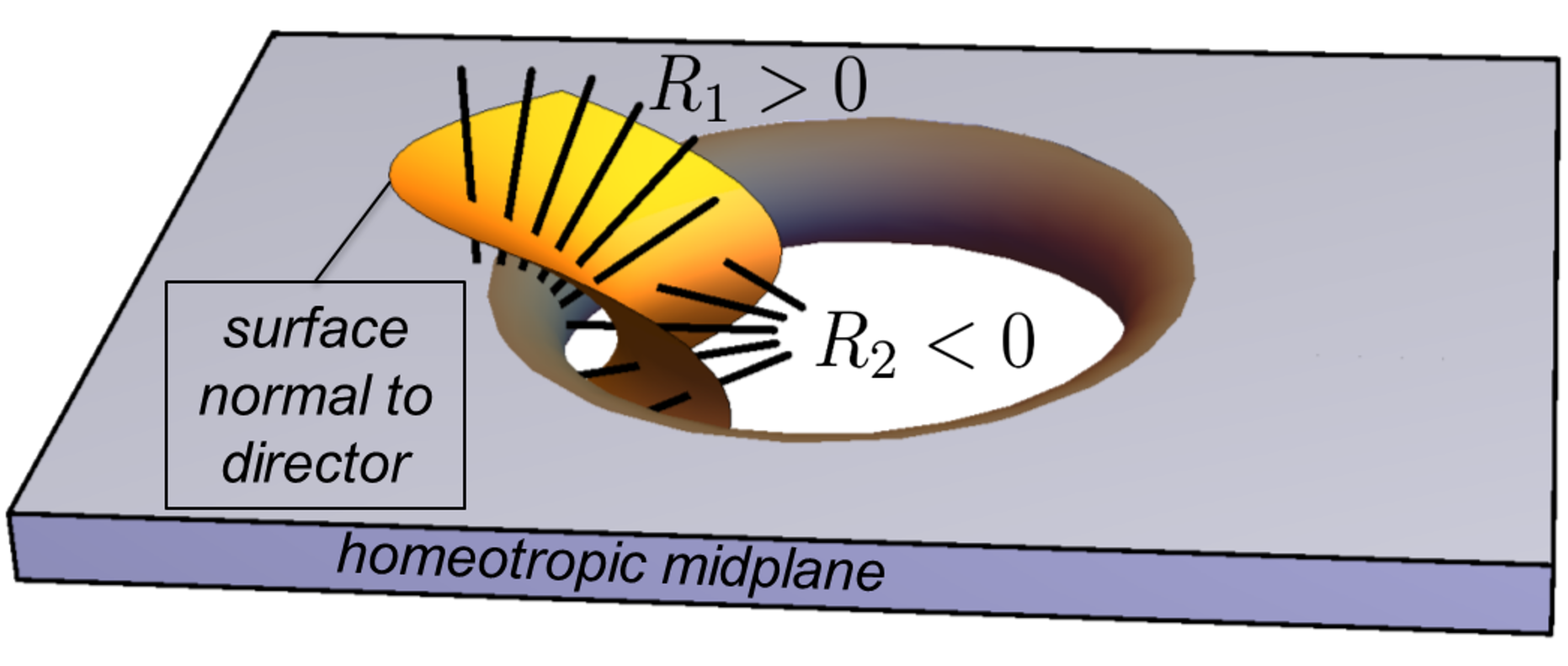}}
\caption{A schematic of a substrate with a hole with homeotropic (perpendicular) anchoring conditions causing a saddle deformation in the bulk. On a hypothetical (yellow) surface,  the boundary conditions along the hole's inner wall favor a  surface normal with a principal radius of curvature $R_2$. When moving from the inner wall  to the top of the substrate, the boundary conditions  favor the normal bending with another principal radius of curvature $R_1$ of opposite sign, indicating that the surface is a saddle. The thick black lines represent the nematic director. \label{FigTheory}}
\end{figure}

\noindent where $\mathbf{n} \equiv \mathbf{n}(\mathbf{x})$ is the (unit) nematic director and $K_1$, $K_2$, and $K_3$ are elastic constants that measure the energy cost for splay, twist, and bend deformations, respectively. The final term with the elastic constant $K_{24}$ is the saddle-splay and, as a total derivative,  is absent from the corresponding Euler-Lagrange equation. However, it contributes to the energy when there are defects, potentially stabilizing them by balancing the energy cost of creating a defect core and the concomitant director distortions  \cite{np-bp-s}.  The saddle-splay term can be rewritten as a surface term through Stokes' theorem, explicitly demonstrating that the saddle-splay is imposed via the boundaries. With strong anchoring of the director at the boundaries, this term therefore offers the possibility of changing the stable or metastable states in the bulk by boundary geometry manipulation.
We may rewrite the saddle-splay in terms of concrete geometric properties of the nematic director. When the director is normal to a surface with principal radii of curvature $R_1$ and $R_2$, the splay and saddle-splay terms in Eq.~\eqref{eq:Frank} are $[\mathbf{n} (\nabla \cdot \mathbf{n})]^2 =  \left[1/R_1+1/R_2\right]^2$ and $ -  \nabla \cdot \left[(\mathbf{n} \cdot \nabla)\mathbf{n}-(\nabla \cdot \mathbf{n})\mathbf{n} \right]= 2/(R_1 R_2)$ \cite{rdkpr},
where the splay energy is proportional to the square of the mean curvature and the saddle-splay energy is proportional to the Gaussian curvature. A saddle deformation in the bulk can be induced if the boundary enforces opposite signs of $R_1$ and $R_2$, that is,  a negative Gaussian curvature.  A positive curvature can not reduce the splay contribution, but we see that negative curvature can -- this is known as the principle of splay cancellation and can stabilize disclinations \cite{pa}.

We develop a boundary that promotes these saddle distortions by creating a thin substrate with a hole removed and homeotropic anchoring on its surface. This is then suspended in the middle of the cell (Fig.~\ref{FigTheory}) (fabrication details to follow). The circular rim of the hole, and the slight rim rounding create principal curvatures of opposite signs, just as the inner half of a torus has negative Gaussian curvature. The anchoring aligns the director  normal to this surface, and the saddle deformation propagates into the NLC bulk. The flat surfaces on the sample top and bottom provide further boundary conditions. When the flat surfaces have homeotropic anchoring, we find configurations with axial symmetry around the hole center. However, oriented (non-degenerate) planar anchoring  breaks the azimuthal symmetry of the hole geometry, which is reflected in the director configurations. We find that a hole \textit{array} causes the distortions from each hole to interact and create complex, but well-defined defect structures. We corroborated our experimental observations with numerical minimization and find that these are, at least, metastable minima.

\section{Homeotropic Anchoring}

\begin{figure}[t]
\centerline{\includegraphics[width=.48\textwidth]{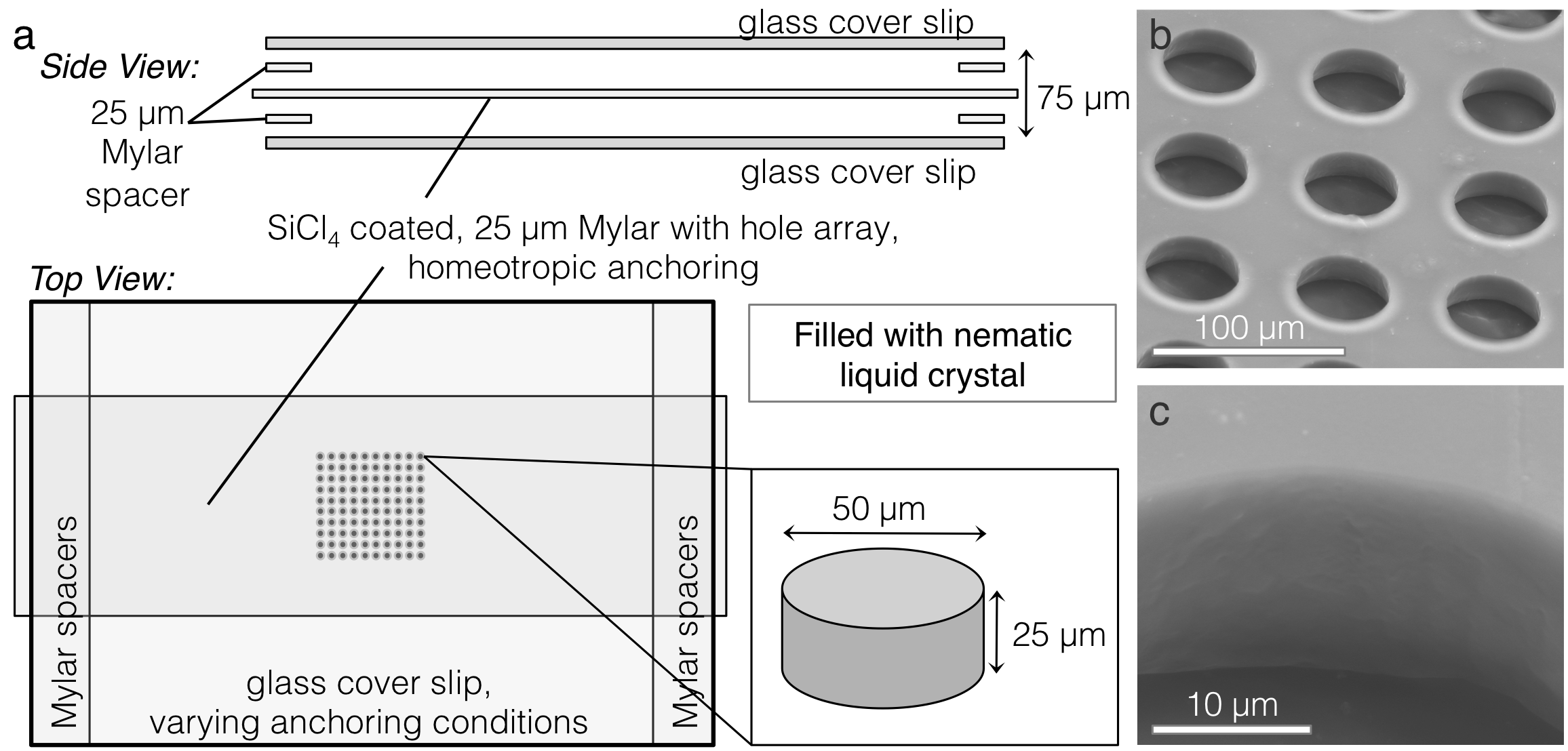}}
\caption{ \label{FigSetup} Experimental setup: (a) Holes with a diameter of 50 $\mu$m are drilled with an excimer laser into a 25-$\mu$m thick Mylar sheet. The sheet is then coated with SiCl${}_4$ to be treated to have homeotropic surface anchoring. The sheet is suspended between two ITO coated glass cover slips with 25 $\mu$m Mylar spacers. These cover slips are treated to have either homeotropic or planar anchoring. (b) and (c): SEM micrographs of the SiCl${}_4$-coated Mylar hole array.  }
\end{figure}

We begin by studying hole arrays in cells with homeotropic anchoring on the top and bottom surfaces, as  illustrated in Fig.~\ref{FigSetup}. A Mylar sheet is used as the hole substrate because of its controlled thickness, smoothness, and transparency, which aids in viewing defects via polarizing microscopy (PM). The LC cell fabrication and assembly  are detailed in Materials and Methods. The LC cells were filled with two types of nematic LC: either the standard, highly birefringent 5CB or a binary mixture of CCN-47 and CCN-55. Separately, at room temperature, the two CCN-compounds are smectic, but their binary mixture is nematic. The CCN mixture is useful for its different elastic constants and its low birefringence, needed for fluorescent confocal polarizing microscopy (FCPM) \cite{fcpm}.  When we anneal our samples, we heat them to the isotropic phase and allow them to cool to the nematic phase, all while a 12 V AC electric field is applied across the sample.  

\begin{figure}
\centerline{\includegraphics[width=.48\textwidth]{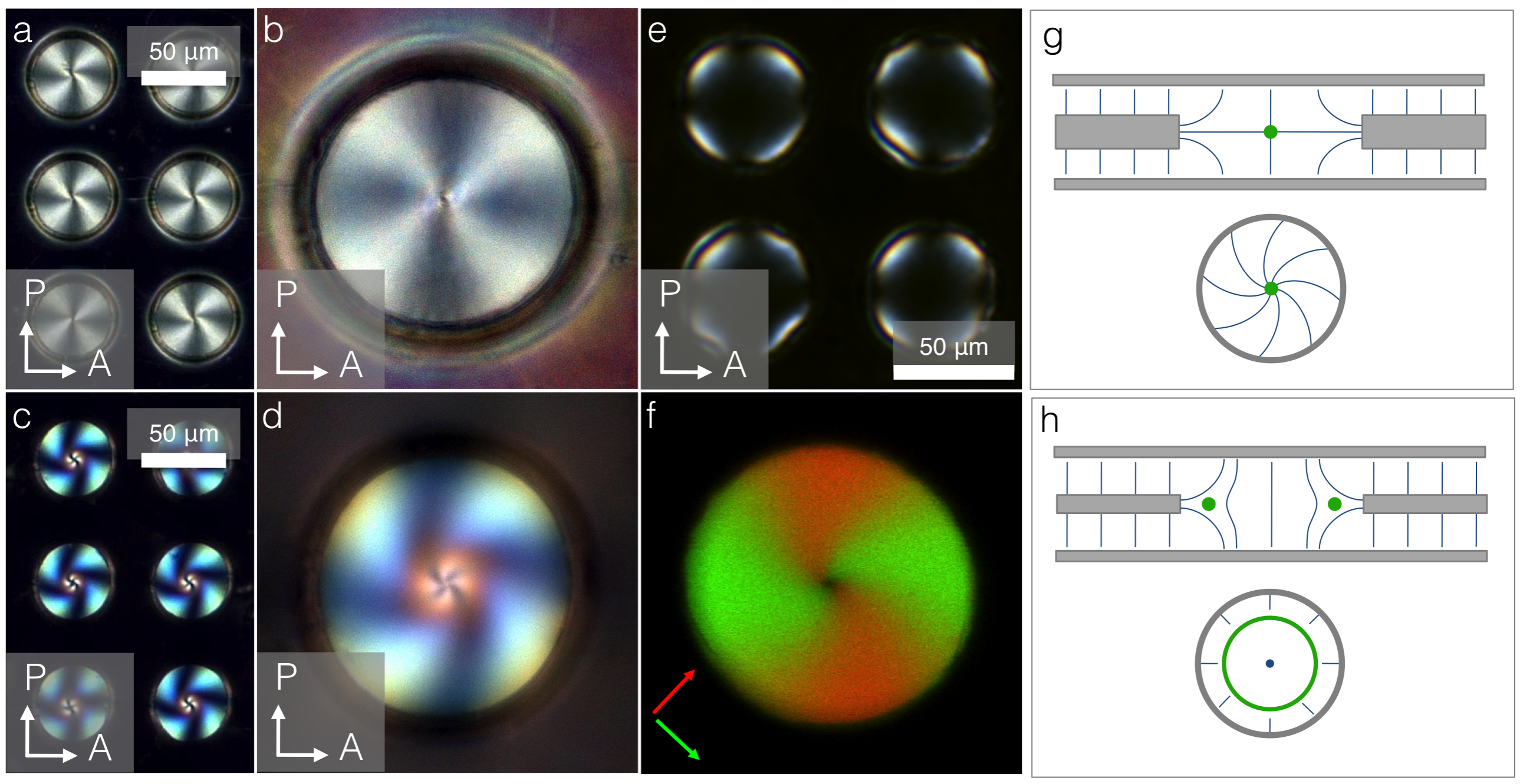}}
\caption{\label{FigHomeo}  (a, b \& e) and (c, d \& f) show the homeotropic system with two different NLCs: 5CB and a binary CCN mixture, respectively. All micrographs were captured via PM, except for (f), which was obtained by overlaying FCPM fluorescent intensities for two perpendicular polarizing directions, indicated by arrows marked in the corresponding color. For a substrate thickness of 25 $\mu$m, point defects are preferred (a, b, c, d \& f), but for a substrate thickness of 6 $\mu$m (e), ring defects occur more often. (g) and (h) show the director configuration for different thicknesses.   }
\end{figure}

Because of the homeotropic anchoring on top and bottom, the net topological charge encoded in the director field vanishes.  Since each hole in the Mylar sheet has a disclination ring that carries hedgehog charge, there must be a companion singularity to satisfy the topological constraint. Based on the geometry, the compensating defect in the LC bulk is expected to have a ``$-1$'' charge, giving either a hyperbolic hedgehog defect or a ring defect with ``$-\nicefrac{1}{2}$'' winding profile, the schematics of which are shown in Fig.~\ref{FigHomeo}(g) and \ref{FigHomeo}(h), respectively.  
At this point, it is useful to recall \cite{rdkpr, mkodl} that though a three-dimensional nematic director, taking values in $\mathbb{R}P^2$, has line defects in three spatial dimensions, these defects do not have a proper winding number as they are classified only by $\pi_1(\mathbb{R}P^2)=\mathbb{Z}_2$. Though one might be tempted to describe the imposed winding at the rim as ``$+\nicefrac{1}{2}$'' with an overall ``+1" charge, that would be incorrect from a topological standpoint.  To make this clear, we will describe this as ``geometric winding'': For example, the rim enforces a geometric winding of $+\nicefrac{1}{2}$.  We warn the reader that geometric winding is not always defined \cite{rdkpr, mkodl} and is, for instance, not defined when the director has a true three-dimensional texture; the twist version of a defect cannot be assigned a geometric winding and, accordingly, switching from regions of $+\nicefrac{1}{2}$ to $-\nicefrac{1}{2}$ geometric winding requires some twist deformation.  However, topology still plays a role: in the presence of the disclination loop around the rim, the $+\nicefrac{1}{2}$ geometric winding necessarily induces a true, topological point hedgehog charge \cite{rdkpr} of the companion defect. The simulated ring state is an example of this and is discussed below.

Indeed, we observe point defects at the hole centers,  as shown through PM and FCPM in Fig.~\ref{FigHomeo}(a-d \& f). The director field around the point defect has a twisted configuration, similar to previously observed defects in nematic droplets with  radial configurations \cite{tw-r-st}. In addition to point defects, sometimes we observe ring defects, also seen in previous work on handlebodies \cite{bs-tc}. When the Mylar sheet is  25 $\mu$m thick, point defects occur significantly more often than ring defects. However, when the thickness  is reduced to 6 $\mu$m, rings appear more frequently, as shown in Fig.~\ref{FigHomeo}(e). This scale-dependence of the defect structure is analogous to the physics of spherical colloids with homeotropic anchoring -- Saturn ring defects become more stable for smaller system sizes compared to those of companion point defects \cite{rvp, ska}.  In our case, a thinner hole substrate has a greater density of splay distortion near the hole rims, favoring the expansion of the central point defect into a ring, allowing the  splay distortion to be canceled near the rim.

\section{Planar Anchoring and Domain Walls}

\begin{figure*}
\centerline{\includegraphics[width=.8\textwidth]{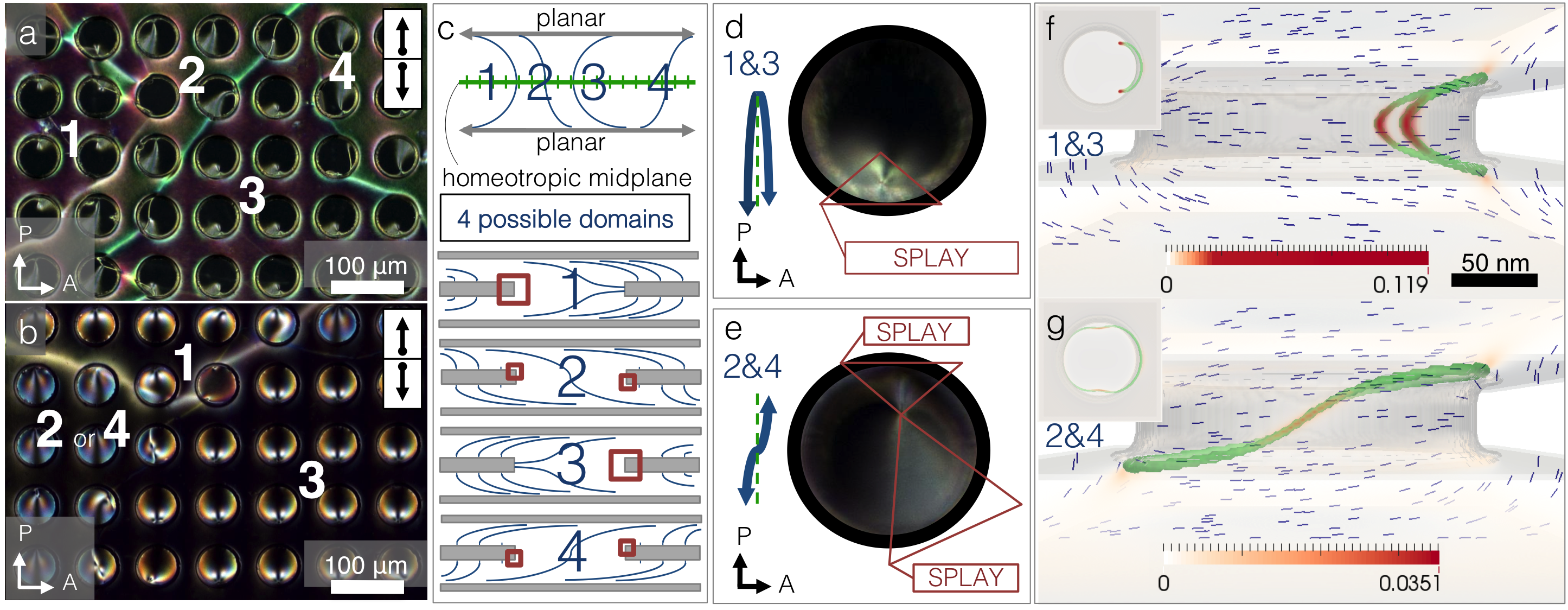}}
\caption{\label{FigPi}  (a) and (b) show the anti-parallel planar system in PM with two different NLCs: 5CB and a binary CCN mixture, respectively. Arrows on the top right represent the glass rubbing direction, with the top box representing the top glass, and likewise for the bottom box and bottom glass. Both 5CB and CCN exhibit domain walls that separate different director configurations within the holes. In (c), the director field can take on either a C or an S formation, resulting in four possible domains. When a hole with homeotropic anchoring is placed into the cell's midplane (c),  certain areas (boxed in red)  near the hole's inner wall have more director field distortion than others for a given configuration. The bend deformations in (c) are also mediated by splay in the holes (d \& e). Numerical results, with splay energy density colormaps (f \& g) (in units of $K_1 /(\Delta x)^2 = 3.3 \times 10^5$ J/m${}^3$, $\Delta x$ being the mesh spacing)  and with the pretilt of the planar substrates included to induce C and S configurations, show ring defects wrapping around the areas of greatest splay (red), in agreement with (c, d, \& e). Defects in the bulk (regions where the order parameter\ $S$ falls to less than 90\% of its equilibrium value: $S<0.9S_0$) are marked in green.}
\end{figure*}

        We investigate configurations that break the hole axial symmetry: anti-parallel-planar ($\pi$-planar) and 90${}^{\circ}$-twisted planar ($\nicefrac{\pi}{2}$-planar), where anti-parallel in experiments refers to the opposite rubbing directions on the top and bottom planar surfaces, coated with polyvinyl alcohol (PVA) (see Materials and Methods). The rubbed PVA does not lie perfectly flat on the surface, but instead the polymer has a slight \textit{pretilt} angle, approximately 1-3$^{\circ}$, in the vertical direction \cite{pretilt, pi-c}, with the angle facing the direction of rubbing.  The equilibrium state of the $\pi$-planar configuration is depicted in Fig.~\ref{FigPi}.  Similar optical textures are seen in  5CB and the CCN mixture (Fig.~\ref{FigPi}(a) and Fig.~\ref{FigPi}(b)).  To understand the  textures, it is useful to consider a  system with the same top and bottom anchoring conditions, but without the perforated Mylar sheet; we can replace the sheet's anchoring conditions with an effective aligning field.  In this case, the physics of the Fr\'eedericksz transition, employed in the traditional twisted-nematic display, should be recalled \cite{meas}.  In  the lower or upper half-cell, the director either ``bends to the left'' or ``bends to the right'' from the midplane to the bottom or top boundary.  This leads to four possibilities shown in Fig.~\ref{FigPi}(c), with two  ``C'' formations and two ``S'' formations.   With perfect planar alignment, all four are degenerate and we see domain walls between them.   The domain walls occur when the curve of the director  changes from bending one way out of the homeotropic midplane ({\sl e.g.} from a C-formation) into bending the other direction ({\sl e.g.} into an S-formation), as shown in Fig.~\ref{FigPi}. In devices, these unwanted domain walls are inhibited through pre-tilting the top and bottom anchoring to bias the bend direction, similarly to our experiments in which the pretilt angle gives rise to a preferred domain after electric field annealing (see Supplementary Video 1).

With this background field structure in mind, we return to the  perforated sheet (bottom of Fig.~\ref{FigPi}(c)). The planar boundary conditions above and below the holes impose more distortion in certain areas of the holes than others, marked in the red boxes in Fig.~\ref{FigPi}(c). These are regions where the larger-scale S or C director curvature is in conflict with the  preferred anchoring direction at the hole rim. In the C-formation, most of the distortion will be located along the rubbing direction axis on the side where the C faces (Fig.~\ref{FigPi}(d)). For the S-formation, the distortion will be along the rubbing direction on both sides (Fig.~\ref{FigPi}(e)). The optical texture asymmetry in Fig.~\ref{FigPi}(e) reflects how the distortion in the S-formation occurs only near the upper or lower hole edge (see  bottom of Fig.~\ref{FigPi}(c)). When the sample is flipped and viewed from the other side, the larger and  smaller bright regions of the optical texture switch locations, showing that the texture asymmetry arises from the distortions' different $z$-locations. Also, near the hole edges, the homeotropic anchoring condition induces a saddle-splay distortion (Fig.~\ref{FigTheory}) that favors splay in  the $xz$ and $xy$ planes, competing with the tendency  to follow the wall anchoring conditions in the $xz$ plane (Fig.~\ref{FigPi}(d) and \ref{FigPi}(e)).

\begin{figure}
\centerline{\includegraphics[width=.48\textwidth]{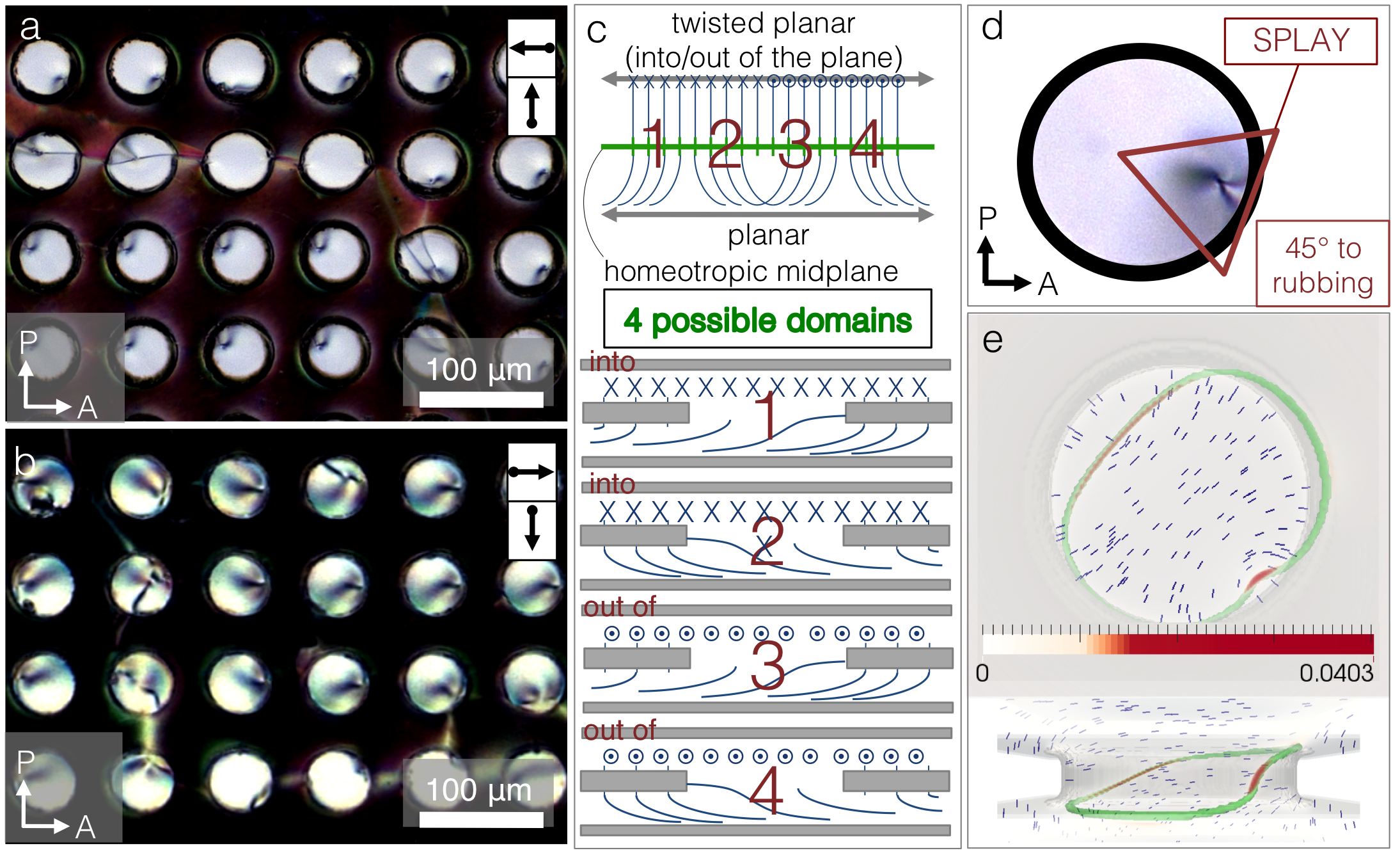}}
\caption{\label{FigPiO2} (a) and (b) show the $\nicefrac{\pi}{2}$-planar cell in PM with two different NLCs: 5CB and a binary CCN mixture, respectively. Arrows on the top right represent the glass rubbing direction, with the top box representing the top glass, and likewise for the bottom box and bottom glass. Both 5CB and CCN show point defects in the holes and domain walls. In (c), the director bends continuously to point in/out to meet the upper planar boundary and left/right to meet the lower planar boundary. When a hole with homeotropic anchoring is placed into the midplane (c), some hole rim areas (marked in (d) by a red triangle) will impose more bend. These areas always occur at $45^\circ$ angles from the rubbing directions (d). Numerical results with a splay energy density colormap (e) (in units of $K_1 /(\Delta x)^2 = 3.3 \times 10^5$ J/m${}^3$, $\Delta x$ being the mesh spacing) show a ring defect wrapping around the hole,  with the greatest distortion  located at $45^{\circ}$ from the rubbing directions, in agreement with (c \& d). Defects are marked in green.}
\end{figure}

The domain walls in the $\nicefrac{\pi}{2}$-planar cell (Fig.~\ref{FigPiO2}) follow the same principle as those in the $\pi$-planar cell. Again, there are four possible domains, seen in experiment (Fig.~\ref{FigPiO2}(a,b,c)). The main difference between the $\pi$ and $\nicefrac{\pi}{2}$-planar cells is the point defect location and the distortion within the hole. Because the distortion must accommodate the director in two different directions above and below the hole, the defect and director distortion will be located in between these two rubbing directions (at 45${}^{\circ}$) (Fig.~\ref{FigPiO2}(d)). As with the $\pi$-cell, the pretilt angles of the planar surfaces pick out the corresponding domain after electric field annealing.

The dark brushes in the optical textures, usually associated with disclination textures in planar systems, do not appear to rotate when the polarizers are rotated. We believe that this is due to the sample thickness and the twist in the director.  The twist might suppress the typical brushes seen in quasi-2D nematic systems.
Our numerical results, described in the next section, show that we do expect  twisted director configurations.  We also checked via numerical minimization ({\it e.g.} Fig.~\ref{FigPi}(f,g) and Fig.~\ref{FigPiO2}(e)) that the   experimental results are consistent with expected equilibrium states.

\section{Numerical Free Energy Minimization}

We employ a  $\mathbf{Q}$-tensor based Landau-de Gennes (LdG) model of a nematic to study the defects in a  cell with the suspended hole array. This model more accurately represents configurations with defects, but reduces to the Frank free energy, Eq.~\ref{eq:Frank}, in the uniaxial limit where the tensor components $Q_{ij}$ are related to the director components $n_i$ via $Q_{ij} = 3S(n_i n_j-\delta_{ij}/3)/2$, where $S$ is the Maier-Saupe order parameter \cite{ms, dgplc}. The LdG free energy was numerically minimized, establishing the director field and the locations of defects \cite{z-rav}. Defect regions are calculated by finding all places where $S<0.9S_0$, with $S_0$ the equilibrium value of the order parameter (see Materials and Methods).  The three eigenvalues of the matrix $Q_{ij}$ may be written as $S$,  and $-S/2\pm S_B$, where $S_B$ is the biaxial order parameter and measures the degree of biaxiality in the system. We found  a maximum ratio $S_B/S_0 \sim 0.1 $ outside of defect regions, with the  majority of values on the order of $10^{-3}$,  justifying our focus on the uniaxial limit.  We used unequal elastic constants that match that of 5CB with a three-constant Landau-de Gennes free energy density. The energy density has an implicit saddle-splay term that is positive and equal to $K_2$ \cite{z-dr,z-k24m}.

The hole array has homeotropic anchoring, and  we set the surfaces 110 nm above and 110 nm below the circular hole array with either planar or homeotropic anchoring.  The hole array itself is also 110 nm thick and the hole rims are rounded with  radius 35 nm. We execute our numerics in a box, periodic in the $xy$-plane, with dimensions $713\times713$ nm${}^2$. The nematic director at each mesh site is oriented in the $z$-direction as an initial condition. We investigate systems for which the oriented directions of planar surfaces are either parallel or rotated 90${}^{\circ}$ relative to one another (twisted planar cell). We  vary the value of $K_2$ (adjusting $L_{24}$ accordingly to keep $K_{24}$ the same) to probe the role of twist deformations on the resulting minimum energy state, as well as the hole diameter to see how geometrical changes alter the defect structures.

\begin{figure}
\centerline{\includegraphics[width=.48\textwidth]{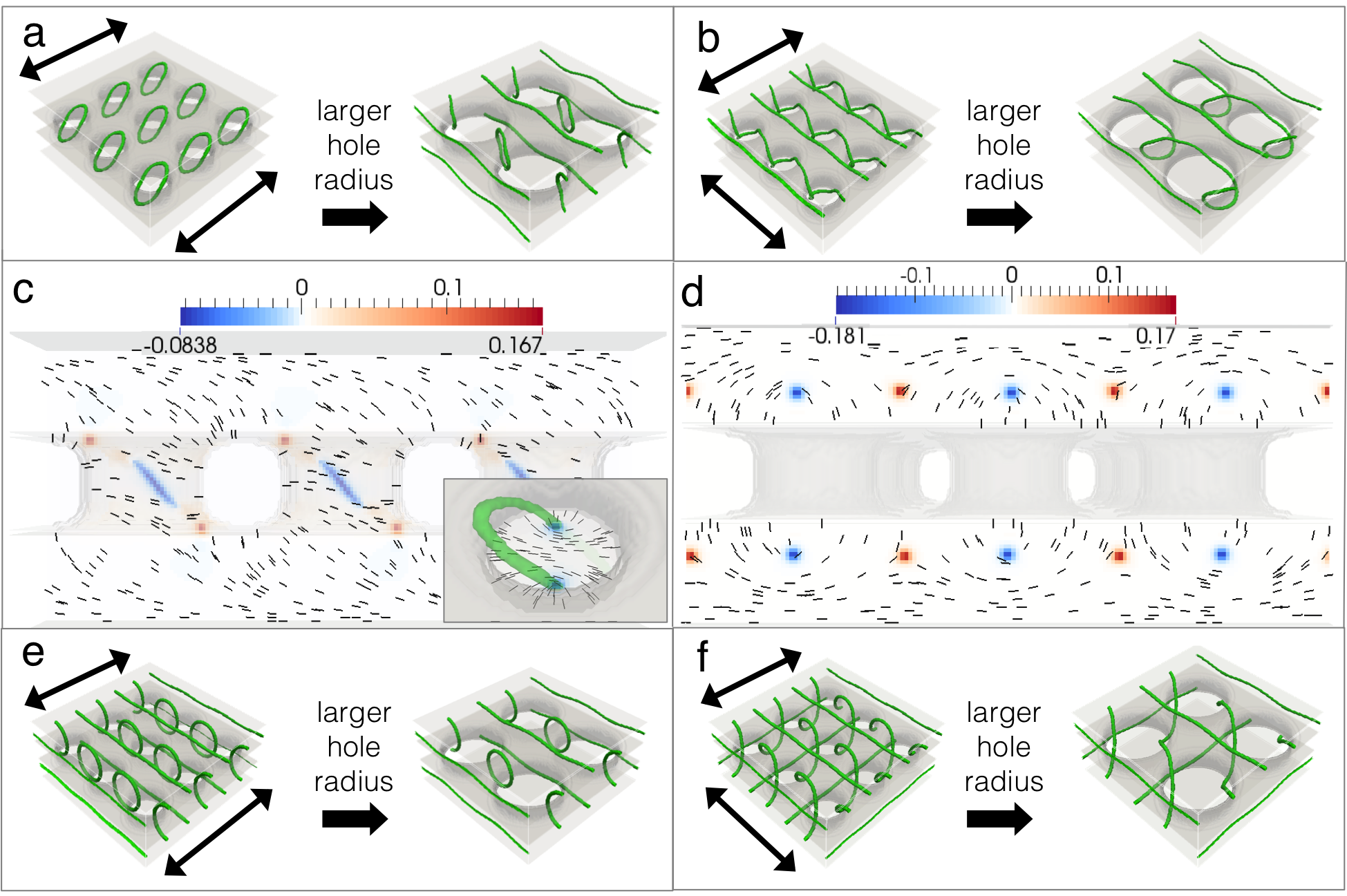}}
\caption{\label{FigSim} A network of disclination lines in a nematic formed with a hole array. Defects are marked in green. The periodic hole substrate has homeotropic anchoring and is suspended between two planar substrates, with arrows indicating the oriented planar anchoring direction. $3 \times 3$  and $2 \times 2$ hole arrays have diameters $d = 132$ nm and $220$ nm, respectively. (a) and (b) use elastic constants matching that of 5CB, and the line state is not stable for smaller diameters in the parallel planar case but is stable for larger diameters in the twisted planar case. (e) and (f) have elastic constants matching that of 5CB, but with a doubled $K_2$ value. The line state is always stable for this case.  Ring state (c) and line state (d) vertical cross sections have saddle-splay energy density colormaps (in units of $|A| = 0.172 \times 10^6$ J/m${}^3$) and demonstrate that positive saddle-splay corresponds to negative geometric winding and vice versa. In the horizontal cross section of the ring (c-inset), areas that have positive geometric winding carry negative saddle-splay.}
\end{figure}

Our numerical results reproduce the observed state with defects localized inside each hole, the so-called \emph{ring state}. However, the simulations also predict a surprising new state:  a \emph{line state} in which disclination lines with geometric $-\nicefrac{1}{2}$ winding form between rows of holes, perpendicular to the rubbing direction of the closest planar surface (Fig.~\ref{FigSim}). In this state, there are additional defects with geometric $+\nicefrac{1}{2}$ winding that wrap around the hole walls. The ordered arrangement of the undulating disclination lines and the defect lines that weave in and out of holes in the twisted planar case (Fig.~\ref{FigSim}(b) and \ref{FigSim}(f)) is reminiscent of those seen in blue phases \cite{bplat, znovbp}.

Though $K_1\sim K_2$ for 5CB, we study the effect of increased $K_2$ in simulations and found that the line state is preferred in the parallel configuration (Fig.~\ref{FigSim}(e)). This suggests that the ring state has a greater amount of twist distortion than the line state.  The geometry of the defect's winding number profile sheds light on the energetic favorability of defect arrangements. For the ring state, cross sections of the ring defect in the $xz$-plane show geometric winding of $-\nicefrac{1}{2}$, while in the $xy$- plane the ring has a geometric winding of $+\nicefrac{1}{2}$, seen in Fig.~\ref{FigSim}(c) through the saddle-splay colormap, where negative saddle-splay corresponds to positive geometric winding and vice versa (explained further below). To switch from one winding  to the other, the nematic director must \emph{twist} and, in this sample geometry, over a short length scale. Similar twisting ring defects were also observed in simulations of highly chiral LCs \cite{ring}. On the other hand, the defects in the line state do not change their geometric winding sign (Fig.~\ref{FigSim}(d)). Cross-sections of the line state reveal that the geometric winding number of the long disclination lines is negative and is positive for the rings  between the holes.  Thus, we expect that when twisting is expensive, the line state will be favored over the ring state.  Further analysis is necessary to determine whether lowering $K_2$ will lead to a stable ring state.

Moreover, we find that saddle-splay distortions help to elucidate the defect structure; compare the saddle-splay energy density of the two states depicted in Fig.~\ref{FigSim}(c) and ~\ref{FigSim}(d).  We plot the saddle-splay density for both the ring state (with 5CB elastic constants) and the line state (with $K_2$ doubled). In both states, regions with positive geometric winding have a negative saddle-splay and {\sl vice versa}. Near the hole we observe that defects with a particular sign of saddle-splay prefer to nucleate near surfaces that induce saddle-splay of the opposite sign (Supplementary Figure 1).  As we varied $K_{24}$ from $-2K_2$ to $2K_2$ in the simulations, the minimum energy state did not change. The saddle-splay distortions, independent of $K_{24}$   in Eq.~\eqref{eq:Frank}, help determine the optimal defect arrangement locally, in agreement with other studies \cite{tord,zsd,2016rav}.

We also find that the ratio of the hole diameter to the inter-hole spacing alters the phases' stability: Larger holes (from 132 nm to 220 nm) stabilize the line state because the larger hole area relative to the intra-hole, homeotropic region increases the influence of the boundary cues on the bulk, establishing director configurations which bend in the directions imposed by the hole edges. Conversely, for smaller holes, the director twists and there are ring defects in the holes\ (Fig.~\ref{FigSim}(c)). The director in the bulk is then free to satisfy the homeotropic anchoring condition between the holes and to uniformly bend in a direction chosen either spontaneously or via the pretilt angle, as it would in a hybrid-anchored cell without holes.

Starting with the initial condition ${\bf n}=\hat z$, a system with parallel (top and bottom) anchoring and with large $K_2$  relaxes to the line state. Note that we can slightly alter these boundary conditions to nucleate the ring state by introducing a small pretilt angle ($3^\circ$ to match that of rubbed PVA \cite{pretilt}) into the $\hat z$-direction.  This increases the energy of the  alternating curving structure of the line state and thus favors the ring state. Alternatively, if we relax the numerics starting from the ring state, the line state never ensues and we find a lower total free energy, suggesting that the line state is metastable and is unstable relative to the ring state under  boundary condition perturbations. Supplementary Video 2 shows how defects in the line state annihilate to make the ring state.

The planar substrate/pretilt angle arrangements also influence the defect locations in simulations. When the planar substrates are parallel, the defect is located on one side of the hole, pinning in areas of highest splay along the hole rim and following a C-formation (Fig.~\ref{FigPi}(f)). When the planar substrates are anti-parallel (the $\pi$-cell), the defect is also pinned on portions of the hole rim edges that have the greatest amount of splay, but on both sides of the hole, following an S-formation (Fig.~\ref{FigPi}(g)). This is consistent with our experimental observations of the defects in C- and S- configurations. We believe that the disparity in defect types, rings in simulations and points in experiments, is due to their large difference in scales (micron-scale for experiments and nano-scale for simulations). 

Let us now return to the line state: is this state we predicted from simulations observable in \emph{experiment}?  

\section{Disclination Line State}

\begin{figure}
\centerline{\includegraphics[width=.47\textwidth]{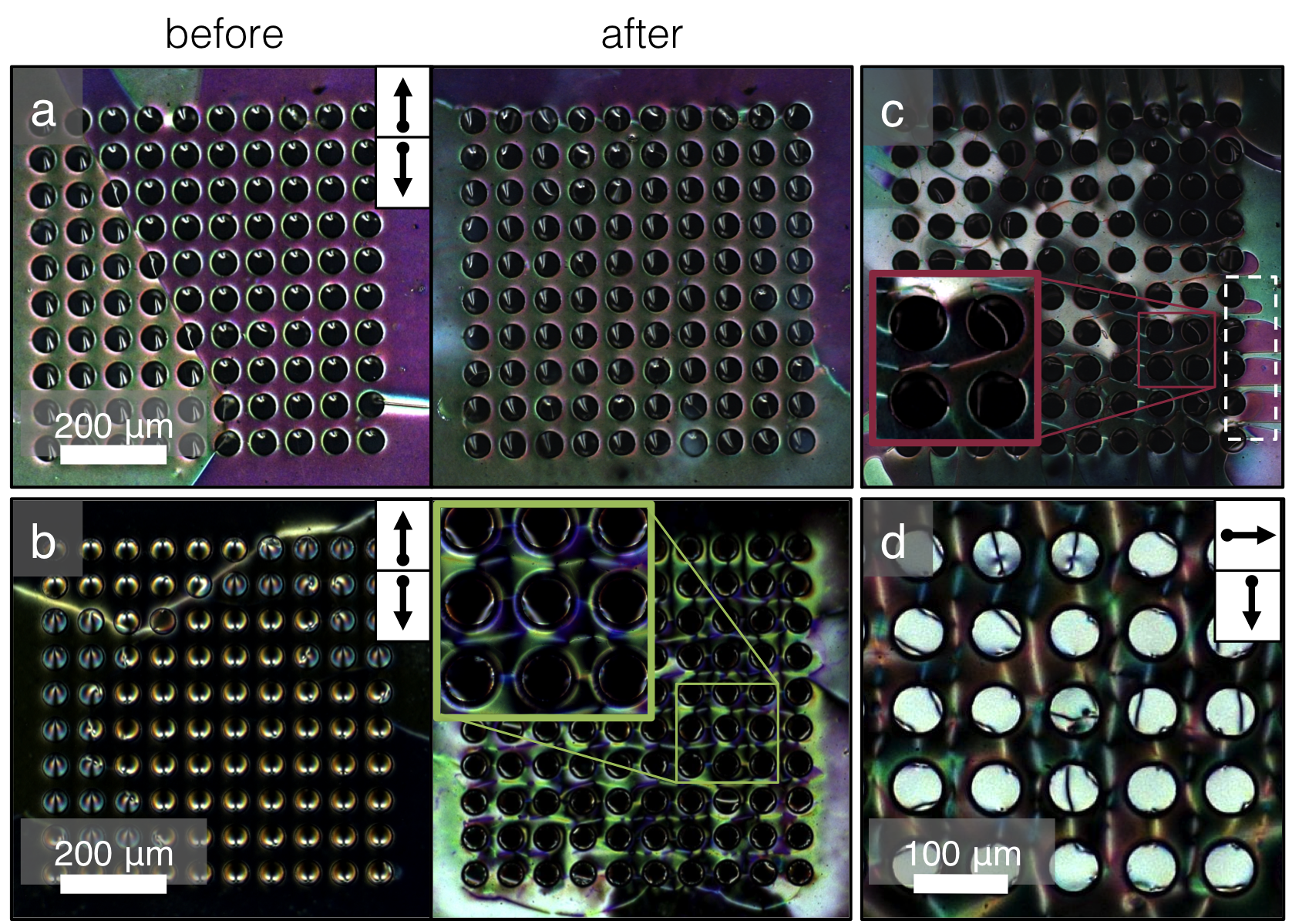}}
\caption{\label{FigLine} Disclination lines (geometric winding $+\nicefrac{1}{2}$) confirmed in experiment with PM. Arrows on the top right represent the glass rubbing direction, with the top box representing the top glass, and likewise for the bottom box and bottom glass. An 12 V AC electric field is applied across a $\pi$-planar cell with a suspended homeotropic hole substrate. The system is then heated and cooled from the isotropic phase back to the nematic phase, after which the field is turned off. For 5CB (a), domain walls across the hole array are annealed away. For the CCN mixture in $\pi$-planar  (b) and $\pi/2$-planar cells (d), undulating disclination lines  running perpendicular to the rubbing direction form between the holes and are stable for over 24 hours. For 5CB, disclination lines form if the phase transition front closes on or near the hole array (c), with or without an applied electric field. The dashed white box highlights the coexistence of the domain walls and the undulating lines in the lines state. Videos of these annealing processes are in Supplementary Materials.}
\end{figure}

For 5CB samples, we can indeed grow the line state! The state appears when the cooling front of the isotropic to nematic phase transition closes on or near the hole array, as shown in Fig.~\ref{FigLine}(c). When this annealing condition is engineered, disclination lines can be seen in the sample, regardless of whether or not an electric field is applied. Otherwise, we do not see the line state, so we conclude that this state is metastable in 5CB, consistent with its low ratio of $K_2/K_1$. However, this state is reproducibly achieved in samples filled with the CCN mixture after annealing with the electric field (Fig.~\ref{FigLine}(b)), (see Supplementary Video 3) suggesting that the lines state is stable for the CCN mixture under these conditions. Here, the lines state can be reliably obtained in CCN samples in both the $\pi$-planar (Fig.~\ref{FigLine}(b)) and $\nicefrac{\pi}{2}$-planar configurations (Fig.~\ref{FigLine}(d)), after which the state persists for over 24 hours. Supplementary videos 3 and 4 show how the system relaxes after electric field annealing. Our numerical results suggest that this line state stability follows from the higher ratio of $K_2/K_1$ in CCN.  There may be other factors, such as different anchoring strengths for CCN and 5CB. The low birefringence makes it difficult to calculate the CCN elastic constants and anchoring strength. Such a calculation would be an interesting focus of future work.

There is a relationship between how the director curves out of the homeotropic mid-plane to meet the planar surface (i.e. what determines the ``four possible domains''),  the planar anchoring strength, and whether or not a domain wall or defect line will form. With 5CB, the planar anchoring is strong: The majority domain (the remaining domain after annealing) is set by the  planar surface arrangement. Any line discontinuity would likely be located near the planar surface to reduce the energy of disobeying the  planar surface anchoring. We  see domain walls form with or without a hole array. On the other hand, for the line state, the director curves out of the  mid-plane in an alternating fashion with a periodicity set by the hole array (Fig.~\ref{FigSim}(d), Fig.~\ref{FigLine}(b)). This alternating curving leads to a discontinuity between rows of holes in the form of a disclination line in the bulk.

To conclude, we can ``lasso up" three-dimensional networks of defect lines in NLCs, along with ordered arrays of point or ring defects, using a perforated sheet with homeotropic anchoring. Even with fixed system topology, a number of distinct equilibrium defect configurations are accessible by varying the boundaries' geometrical parameters. Furthermore, we confirm that the boundary geometry and the geometric winding of defects are correlated; defects with certain saddle-splay distortions arrange near surfaces with oppositely-signed saddle-splay. This principle could be utilized to design surfaces with specific saddle-splay energies to precisely localize defects that have the corresponding geometric winding. The relative ease of inducing defect line networks, all with simple geometric cues, paves the way for more intricate blueprints of self-assembled structures in nematic LCs.

\section*{Materials}

\subsection{Numerical Modeling} We use  a phenomenological LdG free energy of a nematic $\mathbf{Q}$-tensor field, based on the approach reviewed by Ravnik and \v{Z}umer \cite{z-rav, sh2, sh3}. The free energy is minimized in a finite difference scheme on a  cubic mesh, on which a traceless, symmetric rank-2 tensor $\mathbf{Q}$ is defined. The nematic director can be deduced from $\mathbf{Q}$ as the eigenvector that corresponds to the leading eigenvalue $S$. The LdG free energy density is $f_{\mathrm{LdG}}=f_{\mathrm{phase}}+f_{\mathrm{grad}}$, where $f_{\mathrm{phase}}=A Q_{ij} Q_{ji}/2+ BQ_{ij} Q_{jk} Q_{ki}/2+C (Q_{ij} Q_{ji})^2/4$ and $f_{\mathrm{grad}} = L_1 \partial_k Q_{ij} \partial_k Q_{ij}/2+ L_2 \partial_j Q_{ij} \partial_k Q_{ik}/2+ L_3Q_{ij}\partial_i Q_{kl} \partial_j Q_{kl}$, where $\partial_i \equiv \frac{\partial}{\partial x_i}$ and we sum over repeated indices.
In $f_{\mathrm{grad}}$, $L_1= 3.3 \times 10^{-12}$ N, $L_2 = 5.3 \times 10^{-12}$ N, and $L_3 = 3.5 \times 10^{-12}$ N to model 5CB with elastic constants $K_1 = 0.64 \times 10^{-11}$ N, $K_2 = 0.3 \times 10^{-11}$ N, $K_3 = 1 \times 10^{-11}$ N \cite{mkodl}, and $K_{24} = K_2$ in the three-constant approximation. We also take typical values for the material constants of 5CB \cite{z-rav}: $A = -0.172 \times 10^6$ J/m${}^3$, $B = -2.12 \times 10^{6}$ J/m${}^3$, and $C = 1.73\times 10^6$ J/m${}^3$, giving a mesh spacing of 4.4 nm. Defects are where $S<0.9S_0$, with $S_0 \equiv (-B+ \sqrt{B^2 - 24AC}) / 6C \approx 0.533$. The LdG free energy is minimized over $\mathbf{Q}(x)$ using a conjugate gradient algorithm from the ALGLIB package (\url{http://www.alglib.net/}). To model the anchoring, we use a Rapini-Papoular-type surface potential $\Phi_{\mathrm{surf}}=W_0^s \int_s \mathrm{d} A \operatorname{Tr} [(\mathbf{Q}-\mathbf{Q}^s)^2]$, where $Q^s_{ij} = 3S_0(\nu_{i} \nu_j - \delta_{ij}/3)/2$ is the locally preferred $\mathbf{Q}$-tensor at the anchoring surface $s$ ($\nu_i$ is the surface normal for homeotropic or the locally preferred director direction for oriented planar conditions). The potential strengths  are $W_0^s=1\times10^{-2}$ J/m${}^2$  for homeotropic anchoring and $W_0^s=1.5 \times 10^{-5}$ J/m${}^2$ for  oriented planar anchoring, to match the strengths of 5CB on a surface with DMOAP \cite{ska} and rubbed PVA \cite{meas2}, respectively. Energy density colormaps were calculated by computing $f_{\mathrm{grad}}$ and $f_{24} \equiv -L_{24}(\partial_i Q_{ij} \partial_k Q_{jk}-\partial_i Q_{jk} \partial_k Q_{ij})/2$ for a given $\mathbf{Q}(x)$, with altered  constants $L_i$ such that all the   constants $K_i=0$, except for the  component of interest.

\subsection{LCs} We use 5CB (Kingston Chemicals Limited) and a 50/50 mixture of $4'$-butyl-$4$-heptyl-bicyclohexyl-$4$-carbonitrile (CCN-47) and $4$,$4'$-dipentyl-bicyclohexyl-$4$-carbonitrile (CCN-55) (Nematel, GmbH), both thermotropic LCs with a nematic phase at room temperature. 5CB has a positive dielectric constant and the CCN mixture a negative one, resulting in the molecules aligning parallel and perpendicular to the electric field, respectively.

\subsection{Suspended hole array in LC cell} A $10 \times 10$ array of holes with radius of 50 $\mu$m is prepared by repeatedly drilling a Mylar sheet using IPG Microsystem's IX-255 UV excimer laser in the low fluence setting, provided by the University of Pennsylvania's Quattrone Nanofabrication Facility (QNF). The Mylar is coated with silicon tetrachloride (SiCl$_4$) through vapor deposition for the surface to be treated with N,N-dimethl-N-octadecyl-3-aminopropyltrimethoxysilyl (DMOAP; Sigma Aldrich) to obtain strong homeotropic anchoring \cite{ska,sh2,us-1}. Cover slips coated with indium tin oxide (ITO; SPI Supplies), for the application of an electric field across the sample, are treated to have oriented planar anchoring by spin coating a thin layer of polyvinyl alcohol (PVA; Sigma Aldrich), which is subsequently baked at 80${}^{\circ}$C for one hour, then rubbed with a velvet cloth in the desired direction \cite{us-1}. Additional 25 $\mu$m Mylar spacers are used to suspend the hole substrate between the two glass cover slips. An LC droplet was placed on a heated ITO cover slip at 50 ${}^{\circ}$C first. Next, the Mylar spacers are arranged on the cover slip, and then more LC is pipetted onto the hole array before the second cover slip is placed on top. Samples are then clamped and sealed with glue. 

\subsection{Optical Characterization} PM micrographs are taken using an upright microscope in transmission mode furnished with crossed polarizers (Zeiss Axiolmager M1m) and a high-resolution color camera (Zeiss AxioCam HRc). FCPM images are obtained using an inverted IX81 Olympus microscope  with an FV300 Olympus confocal scan box and a half-wave plate between the objective and filter cubes to rotate the scanning laser polarization \cite{fcpm, sh2}. 0.01\% weight of the dye N,N�-Bis(2,5-di-tert-butylphenyl)-3,4,9,10-preylenedicarboximide (BTBP; Sigma Aldrich) was incorporated into the  CCN mixture to allow  LC director determination via FCPM \cite{fcpm, fcpm2}. A scanning laser wavelength of 488 nm was used for dye excitation. The hole array is characterized by environmental scanning electron microscopy (ESEM) on an FEI Quanta 600 FEG ESEM at 10kV, provided by the University of Pennsylvania's Singh Center for Nanotechnology.

\begin{acknowledgments}
We thank O. Lavrentovich, B. Senyuk, Y. Xia, F. Serra, Z. Davidson, and U. Jagodi\v{c} for helpful discussions. We thank T. Baumgart for access to FCPM. We also thank B. Peterson and E. Johnston of the QNF for help with hole array fabrication.  This work was supported by NSF MRSEC Grant DMR11-20901 and NSF DMR12-62047.  D.A.B. was supported by Harvard University through the George F. Carrier Fellowship. R.D.K. was partially supported by a Simons Investigator grant from the Simons Foundation.
\end{acknowledgments}

\bibliography{LassoK24Bib}

\begin{figure*}
\centerline{\includegraphics[width=.8\textwidth]{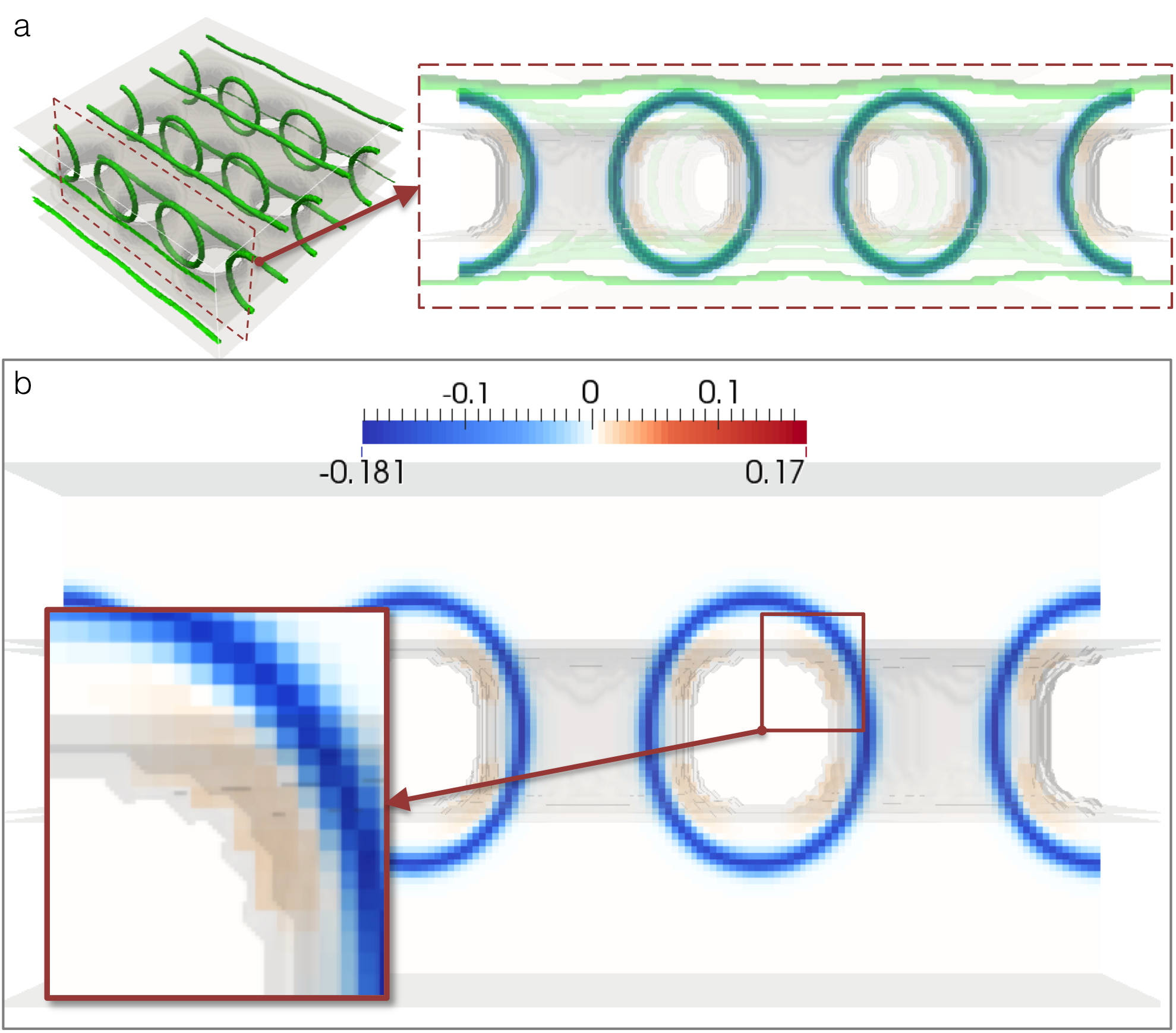}}
\caption{\label{SIFig} Supplementary Figure 1: For the line state (doubled $K_2$), a colormap (in units of $|A| = 0.172x10^{6} J/m_{3}$) (b) of the saddle-splay energy density is made of a cross section (a) of the defects with a positive winding number profile. The positive winding number is associated with a negative saddle splay energy density. A close up of the hole surface shows that the surface has a slightly positive saddle-splay energy contribution. Defects with a negative saddle-splay energy density tend to remain close to surfaces with positive saddle-splay energy density. }
\end{figure*}

\end{document}